\documentclass[onecolumn, superscriptaddress, preprint, 12pt, letterpaper]{revtex4}
\usepackage{amstext}
\usepackage{amsmath}
\usepackage{amssymb}
\usepackage{graphicx}
\usepackage{psfrag}

\newcommand{\ud}{\mathrm{d}}
\newcommand{\vect}[1]{\boldsymbol{#1}}
\newcommand{\eq}[1]{Eq.~\eqref{#1}}
\newcommand{\fig}[1]{Fig.~\ref{#1}}
\newcommand{\stn}[1]{Sec.~\ref{#1}}
\newcommand{\be}{\begin{equation}}
\newcommand{\ee}{\end{equation}}
\newcommand{\ti}[1]{\text{#1}}

\newcommand{\ket}[1]{| #1 \rangle}
\newcommand{\bra}[1]{\langle #1 |}
\newcommand{\w}{\omega}

\begin{document}
\title{Long-lived oscillatory incoherent electron dynamics in molecules:  \textit{trans}-polyacetylene oligomers}
\author{Ignacio Franco}
\email{franco@fhi-berlin.mpg.de}
\affiliation{Theory Department, Fritz Haber Institute of the Max Planck Society, Faradayweg 4-6, 14195 Berlin, Germany.}
\author{Angel Rubio}
\affiliation{Theory Department, Fritz Haber Institute of the Max Planck Society, Faradayweg 4-6, 14195 Berlin, Germany.}
\affiliation{Nano-Bio Spectroscopy group, Dpto. F\'isica de Materiales, Universidad del Pa\'is Vasco, Centro de F\'isica de Materiales CSIC-UPV/EHU-MPC and DIPC, Av. Tolosa 72, E-20018 San Sebasti\'an, Spain}
\author{Paul Brumer}
\affiliation{Chemical Physics Theory Group and Department of Chemistry, University of Toronto, Toronto, Ontario, Canada M5S 3H6}

\date{\today}
\begin{abstract}
We  identify an intriguing feature of the electron-vibrational dynamics of molecular systems via a computational examination of  \emph{trans}-polyacetylene oligomers.  Here,  via the vibronic interactions,  the decay of an electron in the conduction band resonantly excites an electron in the valence band, and vice versa, leading to  oscillatory exchange of electronic population  between two distinct electronic states that lives for up to  tens of picoseconds.  The  oscillatory structure  is reminiscent of beating patterns between quantum states and is strongly suggestive of the presence of  long-lived molecular electronic coherence. Significantly, however, a detailed analysis of the electronic coherence properties shows that the oscillatory structure arises from a purely incoherent process.   These results were obtained by  propagating the coupled dynamics of electronic and vibrational degrees of freedom in a mixed quantum-classical study of the Su-Schrieffer-Heeger Hamiltonian for polyacetylene.  The incoherent process is shown to occur between degenerate electronic states with distinct electronic configurations that are indirectly coupled via a third auxiliary state by the vibronic interactions.  A discussion of how to construct electronic superposition states in molecules that are truly robust to decoherence is also presented. \\
I. Franco, A. Rubio and P. Brumer \emph{New J. Phys.} \textbf{15}, 043004 (2013)
\end{abstract}

\maketitle

\section{Introduction}

A characteristic feature of molecular systems is that they exhibit strong electron-vibrational interactions.  Such vibronic couplings~\cite{Koppel_84,  lukashin75} are an essential component of the photophysics of molecules, leading to vibrations upon electronic excitation~\cite{kobayashi2002}, spectral line broadenings, nonradiative transitions~\cite{Polli2010, martinez2007,christopher05,grinev12}, electronic relaxation~\cite{elsayed99} and decoherence~\cite{Hildner2011, prezhdo_97, hwang04, franco08, joe2011, franco12}.

In this paper we identify an intriguing and novel feature of the electron-vibrational dynamics of \emph{trans}-polyacetylene in which, via the vibronic interactions, the decay of an electron in the conduction band  leads to resonant excitation of an electron in the valence band. The converse process (the decay of an electron in the valence band to a further inner state leading to excitation of an electron in the conduction band) also takes place and brings the system back to its original state.  The result is long-lived oscillatory electron dynamics.   Throughout we  refer to this phenomenon as Vibronically-Induced Resonant Electronic Population Transfer  (VIBRET).

As a model of \emph{trans}-polyacetylene (PA) we employ the Su-Schrieffer-Heeger (SSH) Hamiltonian~\cite{SSH}, a tight-binding model for PA with strong electron-vibrational interactions.  The SSH Hamiltonian is often used to study the static and dynamic features introduced by strong electron-ion couplings in molecular systems~\cite{franco08,stella2011, ness1999}. It has been shown to be successful in capturing the basic electronic structure of PA, its photoinduced vibronic dynamics and the rich photophysics of polarons, breathers and kinks~\cite{kobayashi2009, kobayashi2002, franco08,sergei2003}. The vibronic dynamics of SSH chains is followed by explicitly propagating the coupled dynamics of electronic and vibrational degrees of freedom in an Ehrenfest mixed quantum-classical approximation~\cite{tully, streitwolf1, johansson1, franco08} where the nuclei are treated classically and the electrons quantum mechanically. Effects of nuclear fluctuations and decoherence are captured by propagating an ensemble of trajectories with initial conditions obtained by sampling the ground-state nuclear Wigner phase-space distribution~\cite{wigner}.

Below we show that for a specific class of initial states this model of the vibronic evolution leads to VIBRET in SSH chains that, depending on  system size, can live for up to tens of picoseconds.   VIBRET is seen to arise between degenerate electronic states with distinct electronic configurations that are indirectly coupled via a third auxiliary state by the electron-vibration interactions in the system. Given this  identified level structure, we investigate the effect of changing  system size and the nature of the initial state on the dynamics.

A striking feature of the VIBRET is that it leads to population oscillations among the relevant levels that are analogous to those observed in beatings that result from coherent superposition states. As such, these oscillations seem to indicate that underlying this dynamics  is an electronic superposition state that can live for picoseconds, a timescale that is very long for electronic coherences ~\cite{Hildner2011, hwang04, prezdho06, franco12}.  The question of whether the observed behavior is, in fact, due to a long-lived electronic coherence is  particularly relevant because of  spectroscopic observations in photosynthetic systems that suggest  that unusually long-lived electronic coherences are possible in the Fenna-Matthew-Olson and related complexes~\cite{engel07,collini09, engel_2010}, with timescales exceeding 400-600 fs. Such long-lived electronic coherences have also been noted in intrachain energy migration in conjugated polymers~\cite{scholes2009}.
 Hence, if the SSH model can sustain long-lived coherences even in the presence of strong vibronic couplings, an analysis of the coherence properties of the model may well shed light on this topical problem~\cite{lee07, Mohseni2008, Lloyd2009, fleming09, akihito09, nico2011, coker, kelly2011, pachon_2011, pachon_2012, schulten_2012, fleming_2012,brumer_2012,pachon_2013}.

 We have divided this analysis into three main components. In \stn{stn:vibret} we discuss the essential phenomenology of VIBRET and clarify the basic structure behind the population oscillations. In \stn{stn:coherence} we characterize the coherence properties of VIBRET  by introducing reduced measures of the purity that are apt for many-particle systems. \textit{Using these purity measures we show that, contrary to intuition, the long-lived oscillations observed during VIBRET are the result of an incoherent process.} Last, in \stn{stn:robust} we demonstrate how to construct electronic superpositions in vibronic systems that are truly robust to decoherence. This set of results is expected to have implications in our understanding of vibronic and coherent-like phenomena in molecules, macromolecules and bulk materials.

\section{Model and methods}
\label{stn:model}
In the SSH model,  the PA is described as a tight-binding chain, where each site represents a CH unit, in which the $\pi$-electrons are coupled to distortions in the oligomer backbone by a parametrized electron-vibrational interaction.   For an $N$-membered oligomer, the SSH Hamiltonian reads~\cite{SSH}
 \begin{subequations}
\be
H_\ti{SSH} = H_\ti{e} + H_\ti{ph},
\ee
where
\be
H_\ti{e} = \sum_{n=1, \sigma\pm 1}^{N-1} [-t_0 +\alpha(u_{n+1}-u_n)] (c_{n+1,\sigma}^\dagger c_{n,\sigma} + c_{n,\sigma}^\dagger c_{n+1,\sigma})
\ee
\be
H_\ti{ph} = \sum_{n=1}^{N} \frac{p_n^2}{2M} + \frac{K}{2}\sum_{n=1}^{N-1}(u_{n+1}-u_n)^2,
\ee
\end{subequations}
are, respectively, the electronic ($H_\ti{e}$) and nuclear ($H_\ti{ph}$) parts of the Hamiltonian. Here $u_n$ denotes the displacement of the $n$th CH site from the perfectly periodic position $x=na$ with $a$ the lattice constant of the chain.  $M$ is the mass of the CH group, $p_n$ is the momentum conjugate to $u_n$ and $K$ is an effective spring constant. The operator  $c_{n,\sigma}^\dagger$ (or $c_{n,\sigma}$) creates (or annihilates) a fermion on site $n$ with spin $\sigma$ and satisfies the usual fermionic anticommutation relations. The electronic component of the Hamiltonian consists of a term describing the hopping of $\pi$ electrons along the chain with hopping integral $t_0$ and an electron-ion interaction term with coupling constant $\alpha$.
The quantity $\alpha$ couples the  electronic states to the molecular geometry and constitutes a first-order correction to the hopping integral that depends on the nuclear geometry.   Throughout this work, we assume neutral chains with clamped ends and use the standard set of SSH parameters for PA~\cite{SSH}: $t_0=2.5$ eV, $\alpha=4.1$ eV/\AA, $K=21$ eV/\AA$^2$, $M=1349.14$ eV fs$^2$/\AA$^2$, and $a=1.22$ \AA.   While it is possible to supplement the model with on-site electron-electron interaction terms, for the discussion below these terms are not fundamental and do not change the main findings. We therefore focus on the usual case of noninteracting electrons coupled to phonons.

The method employed to propagate the electron-vibrational dynamics of SSH chains has been described in details previously~\cite{franco08, franco12}. Briefly, the dynamics is followed in the Ehrenfest approximation~\cite{tully},  where the nuclei move classically on a mean-field potential energy surface with forces given by
\be
\label{eq:nuclei}
\dot{p}_n = - \bra{\varphi(t)} \frac{\partial H_\ti{SSH}}{\partial u_n} \ket{\varphi(t)}.
\ee
In turn, the antisymmetrized many-electron wavefunction $\ket{\varphi(t)}$ satisfies the time-dependent Schr\"odinger equation
\be
\label{eq:se}
i\hbar \frac{\partial}{\partial t}\ket{\varphi(t)} = H_\ti{e}[\vect{u}(t)]
\ket{\varphi(t)},
\ee
where $\vect{u}\equiv(u_1, u_2, \cdots, u_N)$.  Decoherence effects are incorporated by propagating an ensemble of quantum-classical trajectories with initial conditions selected by  importance sampling of the ground-state nuclear Wigner distribution function~\cite{franco08, wigner} of the oligomer obtained in the Harmonic approximation.  In this way the dynamics reflects the initial nuclear quantum distribution and is subject to the level broadening and internal relaxation mechanism induced by the vibronic couplings. Results shown here correspond to averages over 10000 trajectories,  providing statistically converged results.

In its minimum energy conformation, the SSH Hamiltonian yields a chain with a perfect alternation of double and single bonds. Its electronic structure is composed of $N/2$ ``valence band" orbitals with negative energies and $N/2$ ``conduction band" orbitals with positive energies that in the long-chain limit are separated by an energy gap of 1.3 eV. The single-particle spectrum depends on the nuclear geometry and changes during the electron-vibrational dynamics. However, because of the electron-hole symmetry in the Hamiltonian, the orbital energies are always such that for each orbital in the valence band of energy $-\epsilon_i$ there is an orbital in the conduction band of energy $\epsilon_i$.

\section{Results and discussion}

\subsection{VIBRET}
\label{stn:vibret}
\subsubsection{Basic phenomenology}

\begin{figure}[tbp]
\centering
\includegraphics[width=0.45\textwidth]{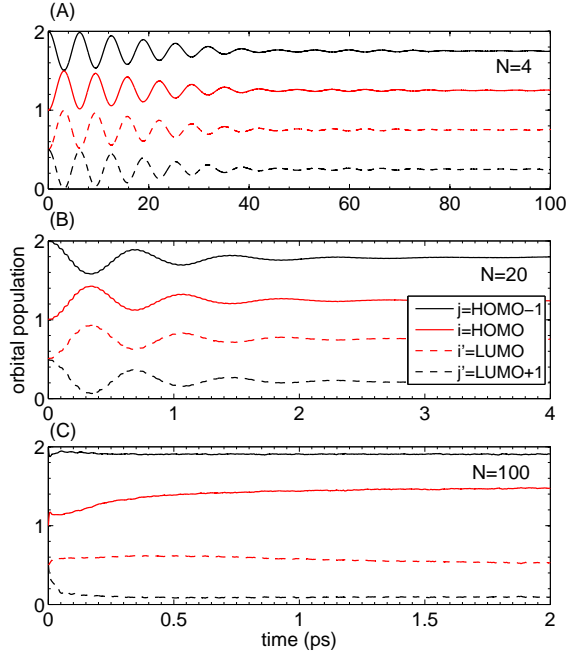}
\caption{VIBRET in  SSH chains with 4, 20 and 100 sites initially prepared in the superposition defined by Eqs.~\eqref{eq:superlonglived}-\eqref{eq:super1}, with $b_0=b_1$. The figure shows the population of the orbitals involved during the complex vibronic evolution. Note the electronic population exchange among levels for $N=4$ and $N=20$.}
     \label{fig1}
\end{figure}

\begin{figure}[tbp]
\centering
\includegraphics[width=0.3\textwidth]{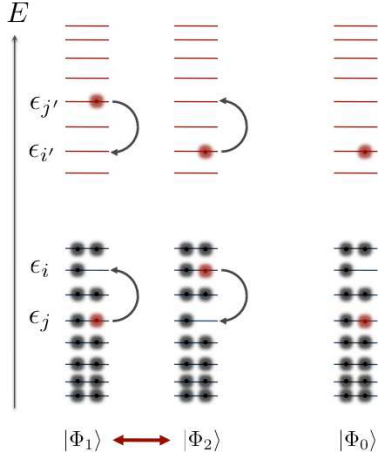}
\caption{Scheme of the single-particle states and the dynamics involved during the VIBRET.  In the figure the horizontal lines denote orbital levels of varying energy $E$ and the circles denote electrons. The arrows indicate the joint electron exchange observed during VIBRET: the decay of an electron from the higher-energy conduction band orbital $j'$ into a lower energy state $i'$ leads to resonant excitation of an electron from an inner state in the valence band $j$ to a higher energy state in the valence band $i$. Upon population inversion, the electron in the valence band $i$ decays back into state $j$ and resonantly excites the electron in level $i'$ to level $j'$. Several of these cycles can be observed when the population exchange is energy conserving, i.e. when $\epsilon_i-\epsilon_j = \epsilon_j'-\epsilon_i'$. The complementary $N$-particle level structure and couplings are depicted in \fig{fig:scheme2}.}
     \label{fig:scheme1}
\end{figure}

\begin{figure}[tbp]
\centering
\includegraphics[width=0.3\textwidth]{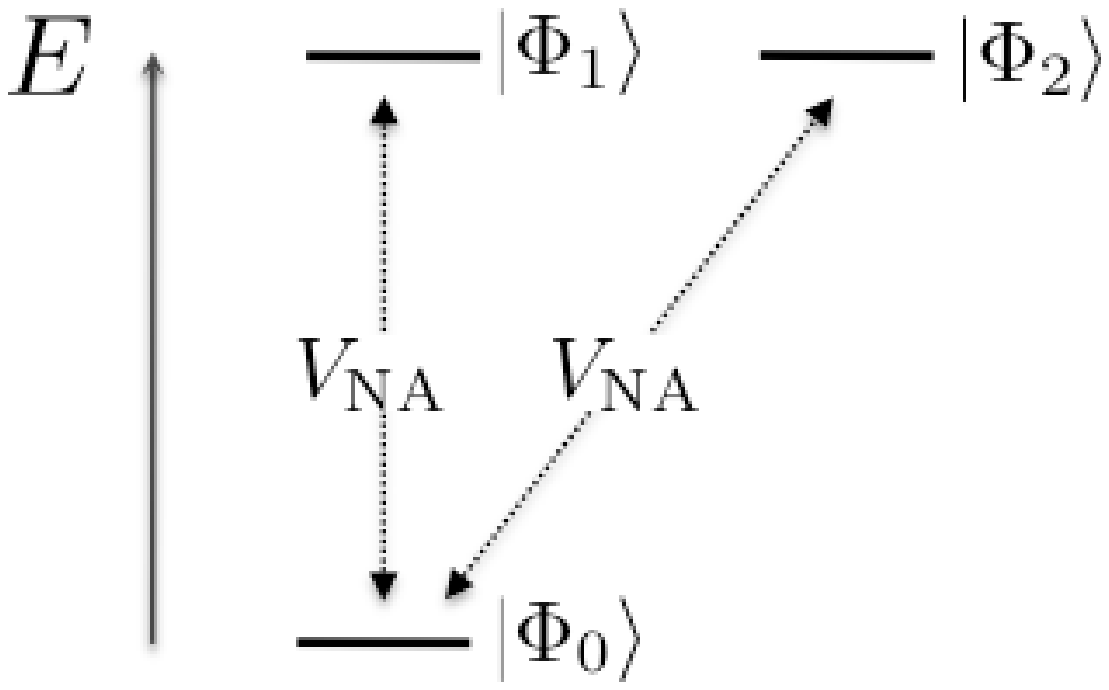}
\caption{Level structure and couplings between $N$-particle electronic states able to sustain VIBRET. The vibronic dynamics couples the  degenerate states  $\ket{\Phi_1}$ and $\ket{\Phi_2}$ indirectly through a third auxiliary state $\ket{\Phi_0}$ via vibronic nonadiabatic coupling terms $V_\textrm{NA}$ [see \eq{eq:nonadia}]  forming a $\Lambda$ or $V$ level system. Figure~\ref{fig:scheme1} depicts one  possible single-particle electronic distribution for the  states in the triad. }
     \label{fig:scheme2}
\end{figure}

We begin by describing the basic phenomenology behind VIBRET. For this, consider the dynamics of an oligomer initially prepared in a separable superposition state of the form
\be
\label{eq:superlonglived}
\ket{\Omega} = (\:b_0\ket{\Phi_{0}} + b_1\ket{\Phi_{1} }\:) \otimes \ket{\chi_{0}}
\ee
where the $\ket{\chi_{0}}$ is the initial nuclear state that, for definitiveness, we take it to be the  ground nuclear state of the ground electronic surface.  As a first example, consider $b_0 = b_1$ and $\ket{\Phi_0}$ and $\ket{\Phi_1}$ to be the states obtained by HOMO$\to$LUMO and HOMO$\to$LUMO+1 transitions from the ground state in a given spin channel.  That is,
\be
\label{eq:super1}
\begin{split}
\ket{\Phi_0} = c_{\textrm{LUMO},\sigma}^\dagger c_{\textrm{HOMO},\sigma}  \ket{\varphi_0} \\
\ket{\Phi_1} = c_{\textrm{LUMO}+1,\sigma}^\dagger c_{\textrm{HOMO},\sigma}  \ket{\varphi_0},
\end{split}
\ee
where $\ket{\varphi_0}$ is the ground electronic state.  Because of the electron-hole symmetry in the SSH approach, state $\ket{\Phi_1}$ is degenerate  with the state
\be
\label{eq:super12}
\ket{\Phi_2} = c_{\textrm{LUMO},\sigma}^\dagger c_{\textrm{HOMO-1},\sigma}  \ket{\varphi_0}.
\ee
States $\ket{\Phi_1}$ and $\ket{\Phi_2}$ have distinct electronic configuration and are thus orthonormal.

Figure \ref{fig1} shows the population dynamics in the four relevant orbitals (HOMO-1, HOMO, LUMO, LUMO+1) involved during the dynamics  for chains of different size ($N=$ 4, 20 and 100). Focus first on  $N=4$ (\fig{fig1}A) where the vibronic dynamics shows a  remarkable  long-lived population transfer between levels that  survive for tens of picoseconds.   Such population exchange is what we refer to as VIBRET.  An analysis of the orbital dynamics leads to the following interpretation of the observed behavior  (a schematic diagram of the population exchange in a generalized setting, and from a single-particle perspective, is shown in \fig{fig:scheme1}).  The population from the higher energy conduction band orbital $j'$ (in \fig{fig1}, $j'=$LUMO+1) is transferred into the lower energy conduction band  orbital $i'$ ($i'=$LUMO in \fig{fig1}). Through the vibronic interactions, this  decay resonantly drives an electron from the inner-most valence orbital $j$ ($j=$HOMO-1 in \fig{fig1}) into the higher energy valence orbital $i$ ($i=$HOMO in \fig{fig1}).  Since the orbital energies are such that $\epsilon_{j'} - \epsilon_{i'} = \epsilon_{i} -\epsilon_j$, the process is energy conserving.   In the case shown in \fig{fig1}A, complete population transfer occurs in $\sim$3 ps.  At this stage, as shown in Fig. \ref{fig:scheme1}, the converse process occurs. Transfer of population from state $i$ to the lower energy state $j$ in the valence band resonantly excites population from $i'$ into $j'$ in the conduction band.  From the perspective of many-particle states, the system is effectively transferring population from state $\ket{\Phi_1}$ to state $\ket{\Phi_2}$, and vice versa. The computed vibronic evolution for the four site chain (\fig{fig1}A) reveals at least six of these $\ket{\Phi_1}\to\ket{\Phi_2}\to\ket{\Phi_1}$ cycles before  the population equilibrates.

We have also observed this intriguing oscillatory population dynamics in 20-site chains (\fig{fig1}B). Here, the period of oscillation  ($\sim0.7$ ps) is faster than in the $N=4$ case and observable for 2 to 3 ps.   That is, while the population dynamics is evident for $N=20$, the process competes with population transfer  into other electronic states that are coupled by the vibronic evolution. Such competition substantially decreases the lifetime of this population exchange with respect to the $N=4$ example, where levels are spectrally isolated. In fact, for $N=100$ (\fig{fig1}C) the  electronic spectrum is so dense that the VIBRET  is simply not observed.

\subsubsection{Level structure and underlying couplings}

A closer look into the three $N$-particle eigenstates  of $H_{e}$   involved in the dynamics, $\ket{\Phi_0}$, $\ket{\Phi_1}$ and $\ket{\Phi_2}$, reveals that they conform to the three-level system schematically shown in \fig{fig:scheme2}. Such system consists of two degenerate orthonormal states, $\ket{\Phi_1}$ and $\ket{\Phi_2}$, with distinct electronic structure that are coupled indirectly through a third eigenstate $\ket{\Phi_0}$ via nonadiabatic coupling terms $V_\textrm{NA}$. The states $\ket{\Phi_1}$ and $\ket{\Phi_2}$ are uncoupled in the vibronic evolution and thus require of the third ``auxiliary" state $\ket{\Phi_0}$ in order to transfer population between one another during the electron-vibrational dynamics. This resonant population transfer is a second order process in $V_\textrm{NA}$ that leaves the population of the auxiliary state approximately constant through the dynamics. 

In order to understand how these effective couplings between levels $\ket{\Phi_0}$, $\ket{\Phi_1}$ and $\ket{\Phi_2}$ arise, consider the selection rules for nonadiabatic couplings in the context of (generic) mixed quantum-classical dynamics. For an electronic wavefunction $\ket{\varphi(t)}$ that satisfies \eq{eq:se} where the time-dependence in $H_e(\vect{u})$ is  assumed to arise from the fact that the nuclei satisfy some trajectory $\vect{u}(t)$, the coefficients in the expansion of $\ket{\varphi(t)}$ in terms of adiabatic eigenstates, i.e. $\ket{\varphi(t)} = \sum_k c_k(t)\ket{\varphi_k[\vect{u}(t)]}$ where $H_e(\vect{u})\ket{\varphi_k[\vect{u}(t)]} = E_k(t)\ket{\varphi_k[\vect{u}(t)] } $, satisfy~\cite{tully_90}:
\be
i\hbar \frac{\ud c_i}{\ud t} = E_i(t) c_i -i \hbar \sum_k \bra{\varphi_i}  \frac{\partial}{\partial t} \ket{\varphi_k} c_k.
\ee
The second term is the nonadiabatic coupling $V_{ik}$ between adiabatic states $i$ and $k$ and can be expressed as:
\be
\label{eq:nonadia}
V_{ik} = -i\hbar\dot{\vect{u}}\cdot \bra{\varphi_i}\nabla_{\vect{u}} \ket{\varphi_k} =  i \hbar\dot{\vect{u}}\cdot \frac{\bra{\varphi_i} \nabla_{\vect{u}}H_e \ket{\varphi_k}}{E_i- E_k}.
\ee
Here we have taken into account the fact that the sole time dependence of the adiabatic states is through the nuclear coordinates such that $\frac{\partial}{\partial t}\ket{\varphi_k} = \dot{\vect{u}} \cdot\nabla_{\vect{u}}\ket{\varphi_k}$.   In SSH chains, $V_{ik}\ne 0$ only if $\bra{\varphi_i} \nabla_{\vect{u}}H_e \ket{\varphi_k}\ne 0$.  This is the case, even for degenerate states. Since $H_e$ is a single particle operator, it then follows that $V_{ik}\ne 0$ between states that differ by at most a single-particle transition. Two particle transitions require terms in the Hamiltonian that are quartic in the creation and annihilation operators, that are absent from this model.  More generally, for systems in which the electron-phonon coupling term in the Hamiltonian is quadratic in the creation and annihilation operators, in order for states to conform to \fig{fig:scheme2} it suffices to guarantee that the selected states $\ket{\Phi_1}$ and $\ket{\Phi_2}$ are degenerate and differ by two particle transitions, but that they are both a single-particle transition away from some state $\ket{\Phi_0}$. This holds even in the presence of electron-electron interactions as they do not contribute to the nonadiabatic transitions. The states employed in \fig{fig1} satisfy precisely these requirements.

\begin{figure}[tbp]
\centering
\includegraphics[width=0.45\textwidth]{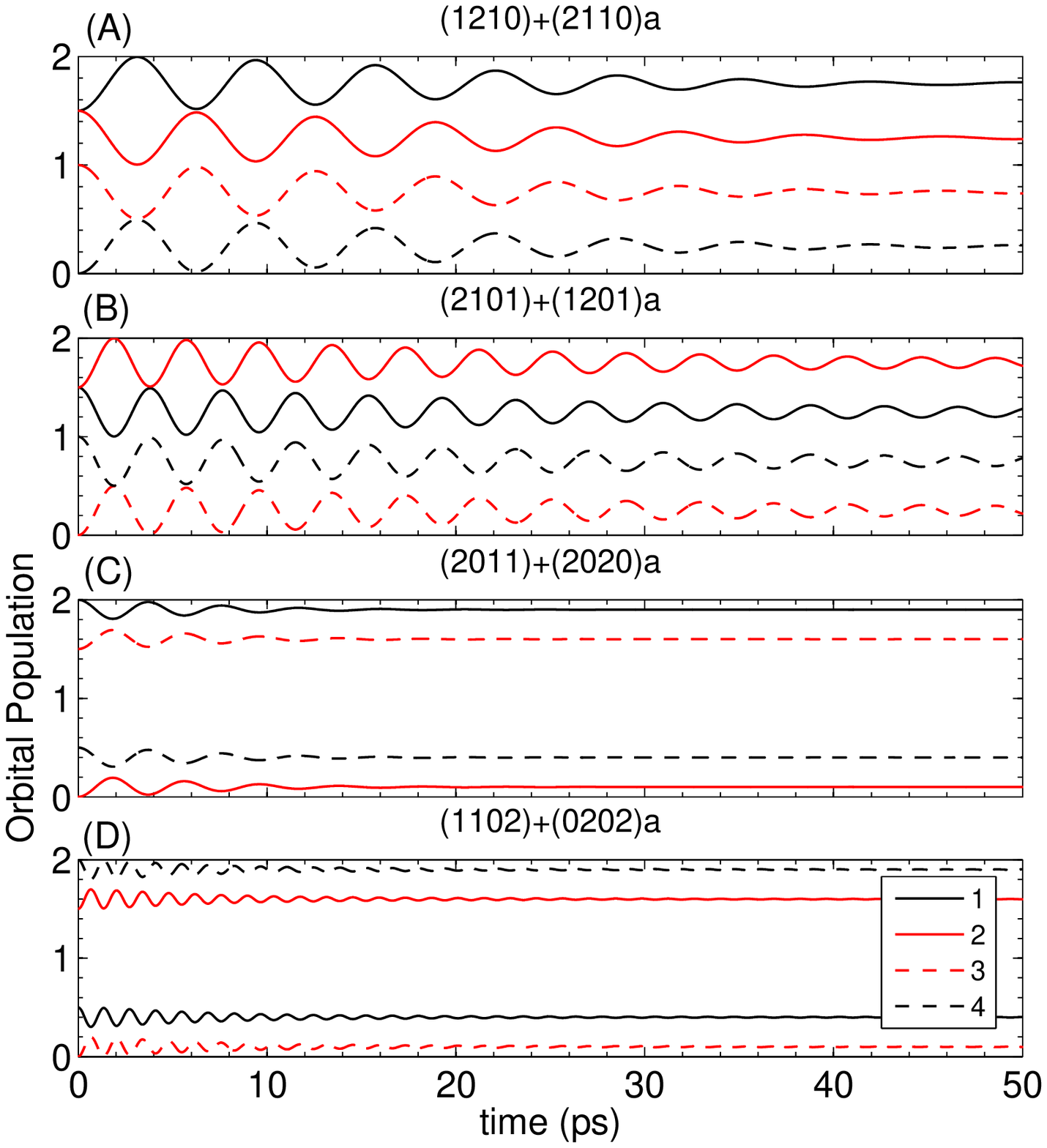}
\caption{Different types of initial electronic superpositions $\ket{\Omega} = (1/\sqrt{2})(\ket{\Phi_0} + \ket{\Phi_1})\otimes\ket{\chi_0}$ for $N=4$ that exhibit the VIBRET. The distribution of the states involved in each case are: (A) $\ket{\Phi_0}= (2110)$,  $\ket{\Phi_1}= (1210)$; (B) $\ket{\Phi_0}= (1201)$,  $\ket{\Phi_1}= (2101)$; (C) $\ket{\Phi_0}= (2020)$,  $\ket{\Phi_1}= (2011)$; (D) $\ket{\Phi_0}= (0202)$,  $\ket{\Phi_1}= (1102)$.  The energies of such states are shown in Table~\ref{tbl:distrib}. The numbers in the legend correspond to the eigenorbital labels in ascending energy. }
     \label{fig2}
\end{figure}

It is now natural to ask if other triads that conform to the scheme  in \fig{fig:scheme2} will also display VIBRET.  For illustrative purposes, we  focus  on the $N=4$ example. In this case there are 19 possible electronic states   and 5 possible degenerate manifolds (without taking into account spin degeneracies); they are tabulated in Table \ref{tbl:distrib}.  The states are labelled by the population of its four  eigenorbitals, in ascending order. In this notation,  the ground state would be state (2200), first excited state (2110), etc. For instance, the superposition employed in \fig{fig1}A would correspond to (2101) + (2110). Figure~\ref{fig2} shows the orbital populations during the dynamics of chains initially prepared in the state of \eq{eq:superlonglived} with $b_0=b_1$ and for different choices of $\ket{\Phi_0}$ and $\ket{\Phi_1}$. The two states involved in the superposition are indicated in each panel. The auxiliary state $\ket{\Phi_0}$  is further labeled by an `a' after the orbital occupations. Figure~\ref{fig2}A corresponds to a situation similar to that described in \fig{fig1}A and \eq{eq:super1}. The auxiliary state $\ket{\Phi_0}$ is the same but now the population, instead of being initially in state (2101),  is initially allocated to the second state in the degenerate manifold (1210). A similar dynamics results, confirming the basic identified level structure.  The dynamics exemplified by \fig{fig2}B also uses the (2101)-(1210) degenerate manifold but here the auxiliary state $\ket{\Phi_0}$ is higher in energy, forming a $\Lambda$ system instead of the $V$ system explored in \fig{fig1}A. Long-lived population transfer between $\ket{\Phi_1}$ and $\ket{\Phi_2}$ is also evident in this case but with a different timescale resulting from the change in the nonadiabatic coupling due to change in $\ket{\Phi_0}$. Figures~\ref{fig2}C and D demonstrate the effect in higher energy degenerate manifolds.  In all cases considered population transfer is as  described in Figs. \ref{fig:scheme1} and~\ref{fig:scheme2},  and  survives for tens of picoseconds. Naturally, if the system is prepared in a state that does not conform to the scheme in \fig{fig:scheme2} no VIBRET will result. Figure S1 exemplifies such a situation for a system initially prepared in state (2200) + (2110).

Note that because the population exchange occurs between degenerate electronic states, there is no net absorption or emission of real phonons during the process. Hence, the limitations of Ehrenfest dynamics in describing spontaneous emission of phonons~\cite{stella2011} do not play a significant role here.

\begin{table}[htbp]
\centering
\begin{tabular}{| c  c   c   c | r | c c c c | r | }
\hline
\multicolumn{4}{|c|}{Distrib.} & E (eV) & \multicolumn{4}{|c|}{Distrib.} & E (eV)\\ \hline
   2 &  2 &  0  & 0 & -11.95   &       1 &  1 &  0  & 2 &  1.81    \\
   \cline{1-5}
      2 &  1 &  1  & 0 & -7.78    &       0 &  2 &  1  & 1 &  1.81    \\
      \cline{1-5}   \cline{6-10}
   1 &  2 &  1  & 0 & -5.97    &    0 &  2 &  0  & 2 &  3.61    \\
   \cline{6-10}
   2 &  1 &  0  & 1 & -5.97    &    1 &  0 &  2  & 1 &  4.17    \\
      \cline{1-5} \cline{6-10}
   1 &  2 &  0  & 1 & -4.17    &    1 &  0 &  1  & 2 &  5.97    \\
   \cline{1-5}
   2 &  0 &  2  & 0 & -3.61    &      0 &  1 &  2  & 1 &  5.97    \\
   \cline{1-5} \cline{6-10}
   1 &  1 &  2  & 0 & -1.81    &        0 &  1 &  1  & 2 &  7.78    \\
   \cline{6-10}
   2 &  0 &  1  & 1 & -1.81    &   0 &  0 &  2  & 2 &  11.95   \\
   \cline{1-5} \cline{6-10}
   2 &  0 &  0  & 2 &  0.00    &     \multicolumn{4}{c}{}\\
   1 &  1 &  1  & 1 &  0.00    &  \multicolumn{4}{c}{} \\
   0 &  2 &  2  & 0 &  0.00   &  \multicolumn{4}{c}{} \\
   \cline{1-5}
\end{tabular}
\caption{Electronic distribution and energy of all 19 possible states for  a $N=4$ PA chain. The distribution correspond to the occupation of the four eigenorbitals in ascending energy. The electronic energy is given at the ground state optimal geometry ($u_1= u_4=0, u_2= 0.0847$ \AA, $u_3= -u_1$). Note the degeneracies in the electronic spectra.}
\label{tbl:distrib}
\end{table}

\begin{figure}[htbp]
\centering
\includegraphics[width=0.45\textwidth]{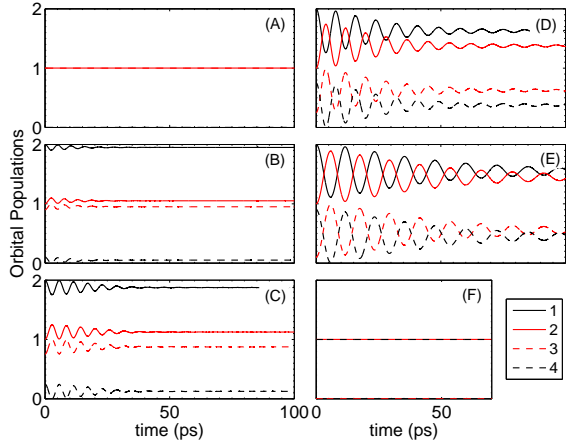}
\caption{Dependence of the VIBRET on the amplitudes of the initial superposition. In this example, the initial state is given by $\ket{\Omega} = (b_0 \ket{\Phi_0} + b_1 \ket{\Phi_1})\ket{\chi_0}$ with $b_0= \sqrt{1- |b_1|^2}$. Results are for $N=4$ and for the superposition defined by the states in  \eq{eq:super1}, i.e. $\ket{\Phi_0}=$ (2110) and $\ket{\Phi_1}=$ (2101).  They correspond to (A) $|b_1|^2= 0$; (B) $|b_1|^2= 0.1$;  (C) $|b_1|^2= 0.25$; (D) $|b_1|^2= 0.75$; (E) $|b_1|^2= 0.9$; (F) $|b_1|^2= 1$. The case of $|b_0|^2=|b_1|^2=0.5$ is shown in \fig{fig1}A. In all cases the initial coefficients where chosen to be real and positive.  The numbers in the legend correspond to the eigenorbital labels in ascending energy.}
     \label{fig3}
\end{figure}

 \subsubsection{Dependence on the amplitudes of the initial superposition}

 Another significant aspect of the observed behavior is the dependence of the VIBRET on the  amplitudes of the states involved in the superposition at the time of preparation. Figure~\ref{fig3} shows the orbital population dynamics for a chain with 4 sites initially prepared in the superposition defined by the states in Eqs.~\eqref{eq:superlonglived}-\eqref{eq:super1}  for different $b_0$ and $b_1$. When all the population is in the auxiliary state (i.e., $|b_0|^2=1$) no population transfer is observed (see \fig{fig3}A) because  the vibrational degrees of freedom are not able to resonantly couple the auxiliary state $\ket{\Phi_0}$  with the degenerate manifold. As the population initially placed in the excited state manifold is increased (the progression shown in \fig{fig3}B-F) the amount of population exchanged during the dynamics changes.  Because of the resonance structure in \fig{fig:scheme2},  only the population that is initially placed in the degenerate manifold can be exchanged, e.g. the state in which initially  $|b_1|^2=0.9$ can exchange at most 0.9 electrons. In addition, by changing the initial coefficients in the superposition one is changing the forces that act on the nuclei at $t=0$ and  thus, effectively, the strength of the nonadiabatic coupling terms during the evolution.   This change in the strength of the nonadiabatic couplings modifies the timescale of the population oscillations and the lifetime of the process. Note that  when all population is placed in the degenerate manifold (\fig{fig3}F) the strength of the nonadiabatic coupling terms is substantially reduced and no population exchange is observed in the simulated time window. We have observed this behavior in all of the cases considered and, as such, it is an inherent feature of this highly nonlinear vibronic evolution. In principle, the phenomenon does not require an initial superposition between states $\ket{\Phi_0}$ and $\ket{\Phi_1}$. In practice, however, such initial coherences introduce additional forces on the nuclei  at initial time that enhance the effective non-adiabatic couplings, leading to a visible effect within the propagated time window.

\subsection{Electronic coherence  during the VIBRET}
\label{stn:coherence}

\begin{figure}[tbp]
\centering
\includegraphics[width=0.45\textwidth]{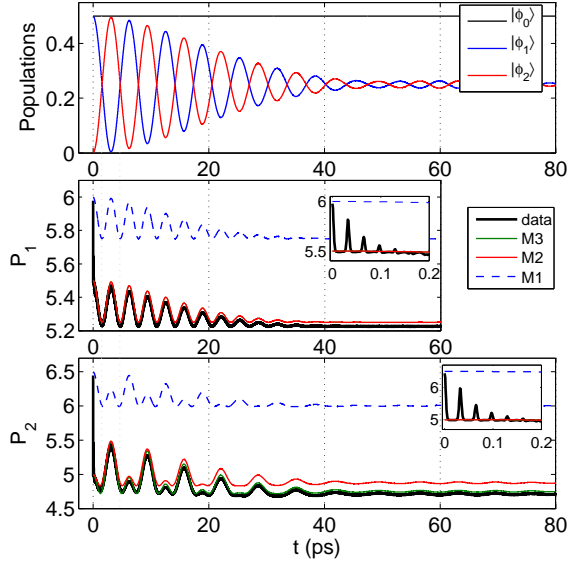}
\caption{Electronic decoherence during VIBRET. The plot shows the populations of the many body states $\ket{\Phi_0}$, $\ket{\Phi_1}$ and $\ket{\Phi_2}$ (top panel), the one-body purity (middle panel) and the two-body purity (bottom panel) for a $N=4$ system initially prepared in the state defined by Eqs.~\eqref{eq:superlonglived}-\eqref{eq:super12} with $b_0=b_1$. The insets highlight the first 200 fs of evolution which are not resolved in the main plots. In the purity plots, the simulated data is shown in black. The colored lines represent a fully coherent (M1), partially coherent (M2) and completely incoherent (M3) model of the state of the system during the dynamics.  In the middle panel the M2 and M3 lines are on top of one another and cannot be distinguished. }
     \label{fig:purities}
\end{figure}

Consider now the electronic coherence properties of the VIBRET in order to determine whether the observed dynamics is coherent or incoherent. For definitiveness we focus on the dynamics in \fig{fig1}A in which the system is initially prepared in the state defined by
Eqs.~\eqref{eq:superlonglived}-\eqref{eq:super12} with $N=4$ and $b_0=b_1$.  The top panel of \fig{fig:purities}  shows the populations of the states $\ket{\Phi_0}$, $\ket{\Phi_1}$ and $\ket{\Phi_2}$, reconstructed  from the orbital populations by supposing that only these three many-particle states participate in the dynamics. The plot clearly shows the population exchange between $\ket{\Phi_1}$ and $\ket{\Phi_2}$, and its decay, while the population of state $\ket{\Phi_0}$ remains approximately constant throughout.

The dynamics exemplified in the top panel of \fig{fig:purities} (also \fig{fig1}A) strongly suggests the presence of a long-lived  electronic coherence because they are reminiscent of beating patterns resulting from superpositions between nearly degenerate states.  If, in fact, the dynamics is a coherent process then the observed evolution would constitute a clear example of a long-lived coherence that is unquestionably electronic.  That is, here, the observed beatings could arise from the effective coupling between states $\ket{\Phi_1}$ and $\ket{\Phi_2}$ that is introduced by the nonadiabatic coupling terms. The decay of the population exchange in the degenerate manifold would then suggest that a decoherence process is taking place with an unusually long decay constant (typical decoherence timescales obtained with this model of the vibronic evolution are of $\sim$10-100 fs~\cite{franco12}).

To examine this possibility we now quantify the coherence properties during VIBRET.
To proceed, it is useful to recall some basic facts about  electronic decoherence in molecular systems. Electronic decoherence in molecules arises because of interactions with the nuclear degrees of freedom and can be understood in terms of nuclear dynamics on alternative electronic potential   energy   surfaces~\cite{hwang04,  prezdho06, prezhdo07, franco08}. To see this, consider the reduced electronic density matrix associated with a general entangled vibronic Born-Oppenheimer state of the form $|\Omega(t) \rangle = \sum_n e^{-iE_n t/\hbar} | \varphi_n \rangle | \chi_n(t) \rangle$,
\be
\begin{split}
\label{eq:rrho}
\rho_e (t) &= \textrm{Tr}_N\{\ket{\Omega(t)}\bra{\Omega(t)}\}\\
&=\sum_{nm} e^{-i\omega_{nm}t} \langle \chi_m (t) | \chi_n (t) \rangle  | \varphi_n\rangle \langle \varphi_m | .
\end{split}
\ee
Here the trace is over the nuclear states, $\ket{\varphi_n}$  are the electronic eigenstates [$H_\ti{elec}\ket{\varphi_n} = E_n \ket{\varphi_n}$], $\ket{\chi_n(t)}$ the nuclear wavepacket associated with each electronic level and $\w_{nm}= (E_n- E_m)/\hbar$.  Note that the magnitude of the off-diagonal elements of $\rho_e(t)$ are proportional to the nuclear overlaps $S_{nm}(t)=\langle \chi_m (t) | \chi_n (t) \rangle $.  Hence, the loss of such  $\rho_e(t)$ coherences is a result of the evolution of the $S_{nm}(t)$ due to the vibronic dynamics.  Standard measures of decoherence capture precisely this.  For example, the purity of such entangled vibronic state \cite{xupei93} is given by
\be
\label{eq:purity}
\textrm{Tr}(\rho_e^2(t)) = \sum_{nm} |\langle \chi_m(t)|\chi_n(t)\rangle |^2
\ee
and decays with the overlaps of the nuclear wavepackets in the different electronic surfaces.

In order to quantify the coherences during VIBRET one ideally would like to study  the purity [\eq{eq:purity}] directly. However, for many-electron systems, like the one considered here, the electronic density matrix  $\hat{\rho_e}$ is a many-body quantity that is not  easy to compute and hence reduced descriptions of the purity are required. Here we introduce and follow the dynamics of the one-body and two-body reduced purities, defined as:
\be
\begin{split}
P_1(t) = \textrm{Tr} \{\hat{\rho}^2(t) \}, \\
P_2(t) = \textrm{Tr} \{\hat{\Gamma}^2(t) \}, \\
\end{split}
\ee
where $\hat{\rho}$ and $\hat{\Gamma}$ refer to the one-body and two-body electronic density matrices. These quantities are defined as:
\be
\rho_p^q = \sum_\sigma \textrm{Tr} \{ c_{p\sigma}^\dagger c_{q\sigma} \hat{\rho}_e\}
\ee
\be
\Gamma_{pq}^{sr} = \frac{1}{2} \sum_{\sigma,\sigma'}  \textrm{Tr} \{ c_{p\sigma}^\dagger c_{q\sigma'}^\dagger c_{r\sigma'}  c_{s\sigma} \hat{\rho}_e\}.
\ee
Because the one-body purity is constructed from the one-body density matrix, it  only informs about coherences between states that differ at most by single-particle transitions. For example, it cannot distinguish between a superposition and a mixture between states that differ by two (or more) particle transitions. Similarly,  the two-body purity is only informative about the coherences between states that differ by at most two particles transitions.

The middle and bottom panels of \fig{fig:purities} shows the dynamics of the reduced purities during VIBRET. In order to interpret the results we consider three models of the state of the system that differ in the assumed degree of coherence. The models are defined by the following  assumed forms for the electronic density matrix:
\be
\label{eq:models}
\begin{split}
\textrm{M1:} \quad \hat{\rho}_e &= (c_0 |\Phi_0\rangle + c_1 | \Phi_1\rangle + c_2 | \Phi_2\rangle) (c_0^\star \langle \Phi_0| + c_1^\star \langle \Phi_1| + c_2^\star \langle \Phi_2|); \\
\textrm{M2:} \quad \hat{\rho}_e &=  |c_0|^2 |\Phi_0\rangle \langle \Phi_0|  + (c_1 | \Phi_1\rangle + c_2 | \Phi_2\rangle) ( c_1^\star \langle \Phi_1| + c_2^\star \langle \Phi_2|) ;\\
\textrm{M3:} \quad \hat{\rho}_e &= |c_0|^2 |\Phi_0\rangle \langle \Phi_0|  + |c_1|^2 | \Phi_1\rangle \langle \Phi_1| + |c_2|^2 | \Phi_2\rangle \langle \Phi_2|.
\end{split}
\ee
They represent, respectively, a fully coherent model (M1), a partially coherent model where only the coherences within the degenerate manifold are maintained (M2), and a fully incoherent model (M3).  The purity resulting from the simulation is shown in black, while the colored lines show the purity expected for these three models, reconstructed by supposing that only the
 $\ket{\Phi_0}$, $\ket{\Phi_1}$ and $\ket{\Phi_2}$  many-particle states participate in the dynamics. The insets highlight the first 200 fs of evolution which are not resolved in the main plots.

From $P_1(t)$ (middle panel) we conclude that a fully coherent picture is not representative of the actual dynamics.  Here, the system begins in a pure state and during the first 200 fs the system displays fast decoherence between the states in the initial  $\ket{\Phi_0}$,$\ket{\Phi_1}$ superposition. The inset details this initial decoherence process. The recurrences observed in the one-body purity signal the vibrational dynamics in the excited state manifold~\cite{franco12} that lead to time-dependence of the overlaps in \eq{eq:purity}. These recurrences are not captured by the models in \eq{eq:models} because they do not take into account  the nuclear  evolution. After this initial fast decoherence, $P_1$ oscillates, reflecting the population changes in the system throughout the dynamics. However, $P_1$ cannot distinguish between the partially coherent model M2 and the fully mixed case M3 because the coherence in M2 is between states that differ by two-particle transitions. In order to distinguish between these two cases we follow the two-body purity $P_2(t)$ shown in the bottom panel. This quantity shows an initial fast decay  (in $\sim$ 200 fs) due to the decoherence between states $\ket{\Phi_0}$ and $\ket{\Phi_1}$, followed by oscillations. After this initial fast decoherence dynamics the model that best adjusts to the observed behavior is M3. That is, the observed population exchange, even when reminiscent of beatings in coherent superpositions, is really best described as a mixed state between $\ket{\Phi_0}$, $\ket{\Phi_1}$ and $\ket{\Phi_2}$. \emph{We thus are forced to conclude that, contrary to intuition, after 200 fs the dynamics during VIBRET is a purely incoherent process. }

The  dynamics is  incoherent  because  superpositions between states $\ket{\Phi_0}$ and $\ket{\Phi_1}$  and $\ket{\Phi_0}$ and $\ket{\Phi_2}$ decohere quickly (on the order of $\sim$200 fs in the $N=4$ case and of 10s of fs for larger oligomers). Since all communication between $\ket{\Phi_1}$ and $\ket{\Phi_2}$ is then through $\ket{\Phi_0}$, a net incoherent process results.  From a quantum-classical perspective, the $V_\textrm{NA}$ in individual trajectories leads to coherences between the states. However, in average the $V_\textrm{NA}$ lead to an incoherent coupling contributing to the incoherent dynamics. The observed directionality in the population exchange is due to the population imbalance between states $\ket{\Phi_1}$ and $\ket{\Phi_2}$. In fact, had we started with a state where $\ket{\Phi_1}$ and $\ket{\Phi_2}$ where equally populated, then no VIBRET would result.

\subsection{What would constitute a superposition of electronic states that is truly robust to decoherence due to vibronic couplings?}
\label{stn:robust}

\begin{figure}[tbp]
\centering
\includegraphics[width=0.4\textwidth]{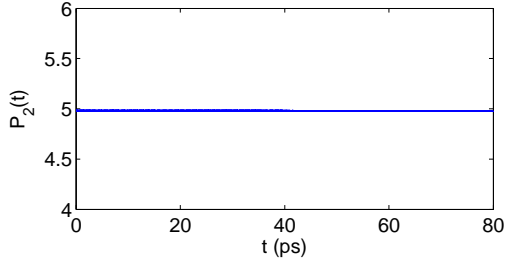}
\caption{Two-body purity for a $N=4$ chain prepared in state \eq{eq:robustsuper}. The quantity $P_2(t)$ shows that this state starts and remains pure during the dynamics even when it is not a stationary state of the Hamiltonian.}
     \label{fig:longlivedcoherence}
\end{figure}

Consider the dynamics where the same vibrational wave packet is prepared in two degenerate electronic states, that is
\be
\label{eq:robustsuper}
\ket{\Omega} = \frac{1}{\sqrt{2}}(\ket{\Phi_1} + \ket{\Phi_2}) \otimes \ket{\chi}.
\ee
Figure~\ref{fig:longlivedcoherence} shows the time-dependence of the two-body purity for a  $N=4$ system prepared in \eq{eq:robustsuper} with $\ket{\Phi_1}$ and $\ket{\Phi_2}$ defined by Eqs.~\eqref{eq:super1} and \eqref{eq:super12}.  For a coherent superposition one expects $P_2(t)$  $\sim$ 5.0 in this case, while a perfectly incoherent state would yield $P_2(t)= 4.5$.  As can be seen,  even when the initial state in \eq{eq:robustsuper} is not an equilibrium state of the vibronic Hamiltonian and leads to a complex electron-vibrational evolution, this initial superposition starts and remains pure throughout the dynamics. It constitutes a clear example of an electronic superposition state with coherence properties that are robust to the vibronic interactions of the chain.

The  feature that  underlies these robust coherences  is the fact that the superposition is between  two electronic states with underlying potential energy surfaces that differ at most by a constant factor.  In this case, the states  $\ket{\Phi_1}$ and $\ket{\Phi_2}$ are degenerate in all conformational space, i.e. $E_1(\vect{u}) = E_2(\vect{u})$ for all $\vect{u}$, where  $E_i(\vect{u})$ is the potential energy surface associated with adiabatic state $i$. Consequently, a given vibrational wavepacket $\ket{\chi}$ will move identically on both surfaces and the nuclear overlap $S_{12}(t)=\langle \chi_1 (t) | \chi_2 (t) \rangle $ that determines the electronic coherence between the two states  [recall \eq{eq:rrho}] is unaffected by the dynamics. In other words,  the two levels involved couple to the environmental bath identically, preventing the bath from entangling with (and thus inducing decoherence of) the system. This is the case provided that the two states are spectrally isolated from other electronic states. 

The quantum structure  involved in these robust coherences falls into the class of decoherence free subspaces~\cite{lidar_98, Kwiat_00}. Such subspace has been suggested (see, e.g., Ref.~\cite{mukamel_11}) to underlie the long coherences in the photosynthetic example, although approximate explicit computations~\cite{alan_2012} have not yet revealed such a structure.
By contrast, the triad in \fig{fig:scheme2} does not  conform to a decoherence free subspace because the potential energy surface of the third electronic state $\ket{\Phi_0}$  generally differs by more than a constant  to that of the degenerate states.


 \section{Conclusions}

 We have identified a new basic feature of the vibronic evolution of a molecular system that we term VIBRET (Vibronically Induced Resonant Electronic Population Transfer). In this process, via the vibronic interactions,  the decay of an electron in the conduction band to a lower energy state resonantly excites an electron in the valence band, and vice versa.  In PA oligomers (as described by the SSH Hamiltonian in a mixed quantum-classical approximation) the population transfer can survive for up to tens of picoseconds and observe several cycles of population exchange.  The process requires two degenerate electronic states with distinct electronic configurations that are indirectly coupled to a third state via vibronic interactions. For Hamiltonians with electron-phonon coupling terms that are at most quadratic in the fermionic operators, such population exchange is realized between degenerate states that differ by two particle transitions but that are both a single-particle transition away from a third auxiliary state.



 The observed population dynamics is strongly suggestive of an electronic coherent process with an unusually long decoherence time. However, things are not always what they seem and, contrary to intuition, an analysis of the one-body and two-body electronic purities shows that  VIBRET occurs incoherently.

We have also demonstrated  electronic superpositions in a molecular system that is robust to decoherence induced by vibronic couplings.  As shown, robust electronic superpositions can arise when the underlying potential energy surfaces of the  states involved in the superposition differ by a constant factor. Under such conditions the vibronic evolution of an initially separable state does not lead to entanglement between the electronic and vibrational degrees of freedom and thus does not lead to decoherence.

We expect the phenomena described here to be of importance in the understanding of vibronic and coherence phenomena in molecules, macromolecules and bulk materials. Future prospects include performing fully quantum simulations of VIBRET and determining ways to manipulate the identified robust electronic coherences.

\begin{acknowledgments}
I.F. thanks the Alexander von Humboldt Foundation for financial support and Dr. Heiko Appel for insightful comments.  A.R. acknowledges financial support from the European Research Council  (ERC-2010-AdG -267374) Spanish Grants (FIS2011-65702-C02-01 and PIB2010US-00652), Grupos Consolidados UPV/EHU (IT-319-07) and EU project  (280879-2 CRONOS CP-FP7). P.B. acknowledges support from the Natural Sciences
 and Engineering Research Council of Canada, and the U. S. Air Force Office of Scientific Research under Contract Number FA9550-10-1-0260. 
\end{acknowledgments}

\bibliography{longlivedpopulations}

\end{document}